\documentclass[11pt]{article}
\usepackage{amstex}
\usepackage{a4wide}

\newif\ifNumberInSections
\newif\ifsubmission

\NumberInSectionstrue

\def\dategerman{\def\today{\number\day.~\ifcase\month\or
  Januar\or Februar\or M\"arz\or April\or Mai\or Juni\or
  Juli\or August\or September\or Oktober\or November\or Dezember\fi
  \space\number\year}}
\def\dateUSenglish{\def\today{\ifcase\month\or
  January\or February\or March\or April\or May\or June\or
  July\or August\or September\or October\or November\or December\fi
  \space\number\day, \number\year}}
\def\dateenglish{\def\today{\number\day \ifcase\day \or
  st\or nd\or rd\or th\or th\or th\or th\or th\or th\or th\or 
  th\or th\or th\or th\or th\or th\or th\or th\or th\or th\or 
  st\or nd\or rd\or th\or th\or th\or th\or th\or th\or th\or 
  st\fi
  ~\ifcase\month\or
  January\or February\or March\or April\or May\or June\or
  July\or August\or September\or October\or November\or December\fi
  \space \number\year}}

\dateenglish

\usepackage{theorem}

\theoremheaderfont{\bfseries\rmfamily\upshape}
		\newtheorem{theorem}{Theorem}
		\newtheorem{lemma}[theorem]{Lemma}

{\theorembodyfont{\rmfamily}

		\newtheorem{acknowledgement}{Acknowledgement}

}

{
\theorembodyfont{\sffamily}
		
}
	
\newtheorem{Scholium}{Scholium}

\ifsubmission
	
	\fi

\ifNumberInSections
\numberwithin{theorem}{section}
\numberwithin{equation}{section}
\numberwithin{figure}{section}
\numberwithin{postulate}{section}
\fi

\newcommand{\qed}{\text{\rule{.4em}{1.7ex}\hspace{.6em}}}
\newenvironment{proof}[1][]{\noindent {\bf Proof#1:\ }}
	{\hspace*{.1em}\hfill\qed\bigskip \noindent}
\newcounter{rom}
\renewcommand{\therom}{(\roman{rom})}
\newenvironment{romanlist}{\begin{list}{\therom}
		{\setlength{\leftmargin}{2em}\usecounter{rom}}}%
{\end{list}}
\newcounter{abc}
\renewcommand{\theabc}{(\alph{abc})}
{\end{list}}

\newcommand{\suchthat}{\bigl |\hspace{.2em}}

\newcommand{\Ric}{{\operatorname{Ric}}}

\renewcommand{\div}{{\operatorname{div}}}
\newcommand{\grad}{{\operatorname{grad}}}
\newcommand{\nab}[2]{\nabla\raisebox{-.8ex}{$#1$}#2}

\newcommand{\p}{\partial}
\renewcommand{\d}{{\mathrm d}}
\newcommand{\D}{{\mathrm D}}

\newcommand{\tr}{{{\mathrm{tr}}}}

\renewcommand{\D}{{\cal D}}

\newcommand{\tx}{{{x}}}
\newcommand{\tg}{{{g}}}
\newcommand{\ttt}{{t}}

\newcommand{\tD}{{\D}}

\newcommand{\hx}{{\hat{x}}}
\newcommand{\hatt}{{\hat{t}}}

\newcommand{\hc}{{\hat{c}}}

\newcommand{\xarray}{{x^1,\dots,x^{m-1}}}

\newcommand{\txarray}{{\tx^1,\dots,\tx^{m-1}}}

\newcommand{\VEC}[2]{\ifnum #2 = 1 
						{#1}^1
					\else 
						{\ifnum  #2 = 2 
							{#1}^1,{#1}^2
						\else
							{\ifnum  #2 = 3 
								{#1}^1,{#1}^2,{#1}^3
							\else
								{#1}^1,\dots,{#1}^{#2}
							\fi}
						\fi}
					\fi}
\newcommand{\xvec}[1]{\VEC{x}#1}
\newcommand{\hxvec}[1]{\VEC{\hx}#1}
\newcommand{\txvec}[1]{\VEC{\tx}#1}

\newcommand{\DRic}{{^{(\D_\nnn)}\Ric}}
\newcommand{\Ds}{{^{(\D_\nnn)}\!s}}

\newcommand{\Dg}{{^{(\D_\nnn)}\!g}}
\newcommand{\Ddg}{\left({^{(\D_\nnn)}\!g}\right)\spdot}

\newcommand{\ds}{{^{(\D)}\!s}}

\newcommand{\dnabla}{{^{(\D)}\nabla}}
\newcommand{\dg}{{^{(\D)}\!g}}

\newcommand{\nnn}{{t}}

\newcommand{\e}{\epsilon}

\newcommand{\MR}{M^+}
\newcommand{\ML}{M^-}
\newcommand{\tttL}{\ttt^-}
\newcommand{\tttR}{\ttt^+}

\title{Distinguished solutions for discontinuous signature change with
weak junction conditions}
\begin{document}
\bibliographystyle{plain}
\author{
Marcus Kriele\thanks{Technische Universit\"at Berlin, Fachbereich
	Mathematik, Sekr. MA 8-3, Stra\ss e des 17. Juni 136, 10623
	Berlin, GERMANY, \protect\newline Email:
	kriele\symbol{"40}sfb288.math.tu-berlin.de, \protect\newline  Telephone: +49 30 314 
	25776, \protect\newline Fax: +49 30 314 21577}}

\maketitle
\ifsubmission\vfill\fi

\begin{abstract}
\noindent
	We consider discontinuous signature change with the weak junction
	condition favoured by Ellis et. al. \cite{ellis-sumeruk+92a}.  We impose
	certain regularity conditions and investigate the space of
	solutions (considered as one-parameter families of
	three-dimensional Riemannian manifolds) for dust and scalar field
	models. 
\end{abstract}


\ifsubmission\vfill\pagebreak\fi

\sloppy

\section{Introduction} 
	Hartle and Hawking \cite{hartle-hawking-83a} have suggested that our universe
	should be described by a signature changing rather than a
	Lorentzian manifold.  Their motivation arose 
	in connection with a Lorentzian path integral approach to quantum
	gravity, and through the hope that in such a setting it may be possible to
	{\em calculate\/} initial conditions for our Lorentzian universe.
	Unfortunately, path integrals in quantum gravity are
	mathematically ill defined.  On the other hand, one can (as a
	semi-classical limit) consider a purely classical theory of
	signature change.  One of the first questions to arises would be 
	the determination of  the space of solutions to  Einstein's equations in the
	presence of signature change.

	The
	qualitative assumption of (classical) signature change may
	constrain the initial data,  and therefore give rise to predictions
	which can be compared with observations of our universe.  {\em As
	far as I can see, this is the only way to test theories of
	signature change.}  Put another way, apart from purely theoretical
	motivations, such predictions are the only way to justify
	(classical) signature change.

	There are several implementations of classical signature change.
	People distinguish between weak and strong junction conditions
	with smooth or discontinuous signature change.  Each of these 4
	flavours represent different conditions one may want to place on
	the transition from Riemannian to Lorentzian signature
	(for a short discussion, see \cite{kriele-95-p-b}).  In this
	paper we are concerned with {\em discontinuous signature and weak
	junction conditions.}  This amounts to solving Einstein's equation
	separately in the Riemannian and the Lorentzian region and then
	matching these two solutions.  This  is different from the strong
	junction condition, i.e., from viewing Einstein's equation
	distributionally and taking into account the non-differentiability
	of the metric at the hypersurface of signature change.

	Discontinuous signature change with weak junction conditions
	recovers (in the analytic category with certain
	$C^{2-}$-conditions on the metric) the whole generality of the
	usual Lorentzian formulation.  Since we are situated in the
	Lorentzian region it follows that (classically) weak signature
	change does not give any physical prediction which can be used to
	differentiate it from the traditional, purely Lorentzian approach.
	It seems that the only way to arrive at predictions from the
	proposal of weak signature change is to impose 
	natural additional   conditions.  It further seems that the only
	way to do so is to demand regularity conditions at the
	hypersurface of signature change on the metric, the energy
	density,  and the principal pressures.

 	In this paper we will investigate a variety of natural conditions.
	In section \ref{s20} we will give a precise definition for
	``discontinuous signature change with weak junction conditions''
	and state our notation.  In section 3 we investigate the effect of
	strong regularity conditions both for dust and scalar field
	spacetimes.  It is shown that Einstein's Equations reduce to a
	highly constrained system of ordinary differential equations.
	While the space of solutions shrinks considerably there exist
	non-trivial solutions.  This will be shown in section \ref{s25}
	where pressureless dust is discussed more systematically.  In this
	chapter we also investigate the effect of (rather weak) regularity
	conditions on the metric and different continuity assumption on
	the energy density.  In section \ref{s50} we discuss the physical
	relevance of our results.
	
\section{Discontinuous signature change and the weak junction condition}
\label{s20}

\noindent
	Different authors give slightly different definitions of
	discontinuous signature change.  We will therefore first state our
	definitions of a ``discontinuously signature type changing
	spacetime $(M,g)$'' (see also \cite{kossowski-kriele-93b}).
	We assume that there exists a hypersurface $\D$ such that
	$M\setminus\D$ consists of two connected components, $\MR$ and
	$\ML$.  $(\MR,g)$ is a  Riemannian and $(\ML,g)$ a 
	Lorentzian manifold.  Moreover, we will demand that both,
	$(\MR,g)$ and $(\ML,g)$ admit  extensions beyond
	$\D$. (But we do not demand a priori that the extension of
	$(\ML,g)$ is in any sense restricted by $(\MR,g)$ or vice versa).
	For points in the Lorentzian region $\ML$ we choose as the fourth
	coordinate, $\tttL$, the Lorentzian distance from $\tD$. Then in
	$\ML$ the metric has the form
\begin{equation*}
-(\d\tttL)^2+\tg^-_{ij}(\tttL,\txarray)\d\tx^i\d\tx^j.	
\end{equation*}
Analogously, we take the negative of the Riemannian distance from
$\tD$ as the fourth coordinate $\tttR$ in the Riemannian region $\MR$.
In $\MR$ the metric is then given by
\begin{equation*}
(\d\tttR)^2+\tg^+_{ij}(\tttR,\txarray)\d\tx^i\d\tx^j.	
\end{equation*}
There are several ways in which to join both regions at $\tD$.
However, it is most natural to demand that $\tg^-_{ij}$ and
$\tg^+_{ij}$ can be joined smoothly\footnote{later in the paper we
will impose different regularity conditions --- cf. {\bfseries Notation and
Conventions}  below}.
Let $\ttt(x) := \tttL(x)$ for $x\in\ML$ and $\ttt(x) := \tttR(x)$ for
$x\in\MR$. We demand that $\ttt$ is a coordinate function on $M$.
Thus we finally obtain
\[
\tg =-\eta \d\ttt^2+\tg_{ij}(\ttt,\txarray)\d\tx^i\d\tx^j,	
\]
where $\eta(x) = 1,\tg_{ij}(x) = \tg_{ij}^-(x)$ for $x\in \ML$ and
$\eta(x) = -1, \tg_{ij}(x) = \tg_{ij}^+(x)$ for $x\in \MR$.

Set $\D_\hatt := \{ x\in M\suchthat \ttt(x) = \hatt\}$.  Then
$(\D_\ttt,\tg_{ij}(\ttt,\txarray))$ can be viewed as a 1-parameter
family of  Riemannian three-dimensional manifolds.

\medskip
\noindent
The strong junction condition would imply that $\p_\ttt g_{ij} = 0$ at
$\ttt = 0$.  The weak junction condition only implies that the family
$\tg_{ij}$ is a $C^1$-1-parameter family of  Riemannian metrics.

\medskip
\noindent
{\bfseries Notation and Conventions:\/} Expressions intrinsic to
$(\tD_\ttt,g_{ij}dx^idx^j)$ carry a superscript $(\D_t)$
(e.g.${}^{(\D_t)}\!g = g_{ij}dx^idx^j$ denotes the induced metric on
the hypersurface $\tD_\ttt$).  All indices run from $1$ to $m-1$,
where $m$ is the dimension of  spacetime.  We
will employ  Einstein's  summation convention.  The curvature scalar is denoted
by $s$.  `$\tr$' is always understood as the trace with respect to the
$(m-1)$-dimensional Riemannian metric $g_{ij}$.  For a bilinear form
$A$ defined on $\D_t$ we write $|A|$ for
$\sqrt{g^{ij}g^{kl}A_{ik}A_{jl}}$.  Differentiation with respect to
$t$ is sometimes denoted by a dot.  The word `smooth' is always
understood as a synonym for `$C^\infty$'.

We will always assume that $\Dg_{ij}$ are analytic functions of
$(\xarray)$.  In all papers we are aware of, this assumption has
(implicitly) been made.  It seems necessary because we are considering
a system of partial differential equations which changes its type.
{\em One cannot expect to find non-analytic solutions in the smooth
category.\/} In particular, for smooth metrics the initial value
problem would not be well posed in the Riemannian region (this is
already apparent for the massless wave equation --- the corresponding
initial value problem has no solutions for non-analytical data).  On
the other hand, such problems are well posed in the analytic category.
In \cite{kossowski-kriele-94b} the authors have shown that the Einstein
equation for smoothly signature changing spacetimes is also well posed
in the analytic category.

We will make different assumptions on the dependence of $\Dg$ on $t$.
In most papers in the field it is assumed that $t \mapsto \Dg(\xarray)$ is
$C^{2-}$.  We will make this assumption in theorems \ref{t25},
\ref{t40}.  However, we will also investigate the strongest possible
regularity condition, i.e. that $t \mapsto \Dg(\xarray)$ is a real analytic map
(cf. theorems \ref{t10}, \ref{t20}, \ref{t22}, \ref{t30}).

\section{The Einstein equation}	\label{s30}

We will study the Einstein equation for pressureless dust, where the
world lines of dust particles intersect $\D$ orthogonally, and a
single, non-interacting scalar field.  Assume that $g$ satisfies

\begin{equation}	\label{e30}
	\Ric-\frac{s}{2}g+\Lambda g = 8\pi \left( T_{\text{\scriptsize
	scalar}}+T_{\text{\scriptsize dust}} \right),
\end{equation}

\noindent
where 

\begin{equation}	\label{e40}
	T_{\text{\scriptsize scalar}} = 8\pi \left(\d\phi\otimes \d\phi
	-\frac{1}{2}\left( g(\grad(\phi),\grad(\phi))
	+V(\phi)\right)g\right)
\end{equation}

\noindent
and

\begin{equation}	\label{e50}
	T_{\text{\scriptsize dust}} =\epsilon\d\ttt\otimes\d\ttt.  
\end{equation}

\noindent
Let $(M,g)$ be a spacetime with discontinuously signature changing
metric $g$ which satisfies Equation (\ref{e30}) in $\ML\cup\MR$.  Then
$(M,g)$ is called a {\em spacetime with an adapted dust-scalar field
model\/}.  If $\phi = 0$ then we call $(M,g)$ a spacetime with an {\em
adapted dust model}.  Clearly, vacuum solutions are special cases.

The equation of motion, $\div(T_{\text{\scriptsize scalar}} +
T_{\text{\scriptsize dust}})= 0$, implies

\begin{equation}	\label{e55}
	0 = 8\pi\left(\Delta\phi-\frac{1}{2}V^\prime(\phi)\right) \d\phi +
	\left(\e \div(\p_\ttt) +\d\e(\p_\ttt) \right) \eta \d\ttt + \e
	g(\nab{\p_\ttt}{\p_\ttt}, \cdot ).
\end{equation}

\noindent
The last summand vanishes since $\p_\ttt$ is a geodesic vector field.
Assume the genericity condition that $\phi$ is not constant in any
open subset of $\D$.  Applying the left hand side of Equation
(\ref{e55}) to any $y\in
T\D_\ttt$ with $\d\phi(y) \neq 0$ we obtain the wave equation

\begin{equation}	\label{e60}
	0 = \Delta\phi-\frac{1}{2}V^\prime(\phi).
\end{equation}

\noindent
Inserting Equation (\ref{e60}) again into Equation (\ref{e55}) we get
the dynamical equation for $\e$,

\begin{equation}	\label{e70}
	\p_\ttt\e =  - \frac{1}{2} \e g^{ij}\p_\ttt g_{ij}.
\end{equation}

\noindent
Observe that these equations have been derived only in $M\setminus\D$.
{\em Again, one has to make a choice how to join these physical fields
across $\D$.}  We will consider different conditions for $\phi$ and $\e$.

\begin{lemma}	\label{l10}
	Let $(M,g)$ be a discontinuously signature changing spacetime with
	an adapted dust-scalar field model.  If $\Dg, \phi$ are $C^{k+2}$
	with respect to $(\ttt,\txarray)$ then at $\D$ we have

\begin{align*}
\begin{split}
	0 &= \left(\p_\ttt\right)^{l} \left(
	\DRic_{ij} - \left(\frac{1}{2}\Ds - \Lambda +
	2\pi g^{kl}\p_{x^k}\phi\p_{x^l}\phi + 2\pi V(\phi)\right)g_{ij} - 4\pi
	\p_{x^i}\phi\p_{x^j}\phi \right)
\end{split}
\\
\begin{split}
	0 &= \left(\p_\ttt\right)^{l} \left(
	{}^{(\D_t)}\!\Delta\phi-\frac{1}{2}V^\prime(\phi)\right)
\end{split}
\end{align*}

\noindent
	for all $0 \leq l \leq k$.
\end{lemma}

\begin{proof}
	The set of partial differential equations for the components
	$g_{ij}$ is analogous to \cite[Equation
	5.4]{kossowski-kriele-94b}\footnote{There is a misprint in the
	quoted set of equations: $m-1$ should be replaced by $m-2$}.  and
	is given by

\begin{equation}	\label{e80}
\p_\ttt\p_\ttt g_{ij} = \eta \left(A_{ij}-
	\frac{1}{m-2} g^{kl}A_{kl}g_{ij}\right), 
\end{equation}

\noindent
	where

\begin{equation}	\label{e90}
\begin{split}
A_{ij} &= -2 \left(\DRic_{ij}-\left(\frac{1}{2}\Ds-
	\Lambda\right)g_{ij}\right)+
\\ &	\quad +\eta \left(g^{kl}\p_\ttt g_{ik}\p_\ttt
	g_{jl}-\frac{1}{2}g^{kl}\p_\ttt g_{kl}\p_\ttt g_{ij} +\frac{1}{4}
	\left(\left(g^{kl}\p_\ttt g_{kl}\right)^2 -3g^{kl}g^{np}\p_\ttt
	g_{kn}\p_\ttt g_{lp}\right) g_{ij} \right)
\\ &    \quad + 8\pi\left(\p_{x^i}\phi\p_{x^j}\phi-\frac{1}{2}\left(-\eta
	\left(\p_t\phi\right)^2+g^{kl}\p_{x^k}\phi\p_{x^l}\phi+
	V(\phi)\right)g_{ij}\right).
\end{split}
\end{equation}

\noindent
	This set of equations has the form

\[
	\p_\ttt\p_\ttt g_{ij} = \eta\left(B_{ij}-
	\frac{1}{m-2}g^{kl}B_{kl}g_{ij}\right) + C_{ij},
\]

\noindent
	where both $B_{ij}$ and $C_{ij}$ depend analytically on $g_{ij},
	\p_\ttt g_{ij}, \p_{x^k}g_{ij}, \p_{x^k}\p_{x^l}g_{ij}, \phi,
	\p_\ttt\phi, \p_{x^k}\phi$.  Thus it follows that
	$\left(\p_\ttt\right)^l \left(B_{ij}-
	\frac{1}{m-2}g^{kl}B_{kl}g_{ij}\right) = 0$ $(l \leq k)$ for any
	$C^{k+2}$-solution.  Taking the trace we obtain
	$\left(\p_\ttt\right)^l g^{kl}B_{kl}g_{ij} = 0$ and therefore also
	$\left(\p_\ttt\right)^l B_{ij} = 0$ which proves the first
	assertion of the lemma.

	Equation (\ref{e60}) can be written as $0 =
	{}^{(\D_t)}\!\Delta\phi-\frac{1}{2}V^\prime(\phi) -
	\eta\left(\p_\ttt\p_\ttt\phi + \frac{1}{2} g^{kl}\p_\ttt g_{kl}
	\p_\ttt\phi\right)$.  Thus the second assertion follows by an
	analogous argument.
\end{proof}

\begin{theorem} \label{t10}
	Let $(M,g)$ be a discontinuously signature changing spacetime
	with an adapted dust-scalar field model.  If $\Dg, \phi$ are real
	analytic with respect to $(\ttt,\txarray)$, then the dynamical
	part of the Einstein equation reduces to the system of ordinary
	differential equations

\begin{align*}
\begin{split}
	\p_\ttt\p_\ttt g_{ij} &= C_{ij}- \frac{1}{m-2} g^{kl}C_{kl}g_{ij},
\end{split}
\\
\begin{split}
	\p_\ttt\p_\ttt \phi &= -\frac{1}{2}g^{ij}\p_\ttt g_{ij} \p_\ttt
	\phi
\end{split}
\\
\begin{split}
	\p_\ttt\epsilon 	&=	-\frac{1}{2}\epsilon\tg^{kl}\p_\ttt g_{kl},
\end{split}
\end{align*}

\noindent
	where $C_{ij} = g^{kl}\p_\ttt g_{ik}\p_\ttt
	g_{jl}-\frac{1}{2}g^{kl}\p_\ttt g_{kl}\p_\ttt g_{ij} +\frac{1}{4}
	\left(\left(g^{kl}\p_\ttt g_{kl}\right)^2 -3g^{kl}g^{np}\p_\ttt
	g_{kn}\p_\ttt g_{lp}\right) g_{ij}
	+4\pi\left(\p_t\phi\right)^2$.

\noindent
	In addition, in each surface $\D_\ttt$ the intrinsic equations

\begin{align*}
\begin{split}
	0 &= {}^{(\D_t)}\!\Delta\phi-\frac{1}{2}V^\prime(\phi)
\end{split}
\\
\begin{split}
	0 &= \DRic_{ij} - \left(\frac{1}{2}\Ds - \Lambda +
	2g^{kl}\p_{x^k}\phi\p_{x^l}\phi + 2V(\phi)\right)g_{ij} - 4\pi
	\p_{x^i}\phi\p_{x^j}\phi
\end{split}
\end{align*}

\noindent
	are satisfied.
\end{theorem}

\begin{proof}
Since $g_{ij}, \phi$ are analytic lemma \ref{l10} implies $B_{ij} = 0$
and $0 = {}^{(\D_t)}\!\Delta\phi-\frac{1}{2}V^\prime(\phi)$ at all
hypersurfaces $\D_t$.  The theorem follows by inserting these
equations into the  Einstein equations for $g_{ij}$ and Equation
(\ref{e60}). 
\end{proof}
 
\noindent
	We see that the system is strongly over-determined and it should
	not come as a surprise that there exist only very few solutions.

\section{Pure, pressureless dust $\phi = 0$} \label{s25}

	In order to arrive at concrete results, we will consider one of
	the simplest matter models, pure, pressureless dust.  The only
	matter quantity is the energy density $\e$.  Our main physical
	assumption will be a continuity assumption on $\e$.  There are two
	possibilities which seem to be especially natural.  One may assume
	that $\e$ is a continuous function.  This is carried out in
	section \ref{ss25.10}.  However, in the definition of energy, time
	enters in a fundamental way.  Hence it seems also plausible to
	expect that the change of signature is reflected in the energy
	density by a change of sign.  In subsection \ref{ss25.20} we will
	therefore investigate the assumption that $\eta \e$ is continuous.

	We will consider the two extreme regularity conditions on the map
	$t \mapsto \Dg(\xarray)$, namely that $\Dg$ is an analytic function of $t$
	and that it is merely a $C^{2-}$-function of $t$.

\subsection{Continuous energy density $\e$}\label{ss25.10} 

\begin{theorem}\label{t20}
	Let $(M,g)$ be a 4-dimensional discontinuously signature changing
	spacetime with adapted dust model. If $\Dg$ is real analytic and
	$\e$ is continuous with respect to $(\ttt,\txvec3)$ then 
	$(M,g)$ has the following properties:

\begin{romanlist}
\item\label{t20i}
	$(M,g)$ is a vacuum spacetime: $\e = 0$;
\item	\label{t20v}
	The submanifolds $(\D_t,\Dg)$ are flat;
\item	\label{t20ii}
	The eigenspaces of $g^{ij}\dot g_{jk}$ are constant with respect
	to $t$ (here we are identifying the manifolds $\D_t$ via projection
	along the $t$-coordinate);
\item	\label{t20iii}
	The eigenvalues of $g^{ij}\dot g_{jk}$ are given by $c_i  =
	\frac{2\hc_i}{2+ t \sum_{k=1}^3 \hc_k}$, where
$\hc_i(\xvec3) = c_i(0,\xvec{3})$; 
\item	\label{t20iv}
	$c_1c_2+c_2c_3+c_3c_1 = 0$.
\end{romanlist}
\end{theorem}

\begin{proof}
	It follows from the additional intrinsic equations in theorem \ref{t10}
	that the hypersurfaces $\tD_\ttt$ must be Einstein
	manifolds: $\DRic_{ij} = \left(\frac{1}{2}\Ds - \Lambda
	\right)g_{ij}$.  Taking the trace it follows immediately that
	$\DRic = {2\Lambda}\Dg$.  In particular, $\DRic$ does
	not depend on $\ttt$.
\begin{align}	\label{e95}
\begin{split}
	8\pi\epsilon &= G_{\ttt\ttt} + \Lambda(-\eta)
=	\frac{\eta}{2}\,\Ds+\frac{1}{8}\left(\left(\tr(\left(
	\Dg\right)\spdot)\right)^2 -\left|\left(
	\Dg\right)\spdot\right|^2\right) -\eta \Lambda
\\ &=  {2\eta}\Lambda + \frac{1}{8}\left(\left(\tr(\left(
	\Dg\right)\spdot)\right)^2 -\left|\left(
	\Dg\right)\spdot\right|^2\right)
\end{split}
\end{align}

\noindent
	implies $\Lambda = 0$ by the continuity of $\e$.  But
	this means $\DRic = 0$.  Since for 3-dimensional manifolds the
	Ricci tensor already determines the Riemann tensor all surfaces $t
	= \text{const.}$ must be flat (hence \ref{t20v} follows).

	Consider  a coordinate system of $\D$ such that at a given
	point $(\hxvec3)$ $g_{ij}$ and $\dot g_{ij}$ are simultaneously
	diagonal.  (Such a coordinate system always exist since $g_{ij}$
	is positive definite and $\dot g_{ij}$ is symmetric).  Then
	$g_{ij}$ and $\dot g_{ij}$ are diagonal at $(\ttt,\hxvec3)$ for
	all $\ttt$.  This follows from the uniqueness of solutions for
	ordinary differential equations and the fact that the ansatz

\begin{equation*}
	g_{ij} = 
\begin{pmatrix}
	a_1(t) & 0 & 0
\\
	0 & a_2(t) & 0 
\\
	0 & 0 & a_3(t)
\end{pmatrix},
	\qquad\dot g_{ij} = 
\begin{pmatrix}
	b_1(t) & 0 & 0
\\
	0 & b_2(t) & 0 
\\
	0 & 0 & b_3(t)
\end{pmatrix} \ 
\end{equation*}

\noindent
	leads to a consistent system of differential equations. In fact,
	one obtains

\begin{equation*}
	C_{ij} = \sum_{k=1}^3 \frac{1}{a_k} b_i b_j \delta_{ik}\delta_{jk}
	-\frac{1}{2} \sum_{k=1}^3 \frac{b_k}{a_k} b_i \delta_{ij} +
	\frac{1}{4}\left( \left(\sum_{k=1}^3 \frac{b_k}{a_k}\right)^2 - 
	3\sum_{k=1}^3 \left(\frac{b_k}{a_k}\right)^2\right) a_i
	\delta_{ij}
\end{equation*}

\noindent
	(no summation over $i, j$) whence $C_{ij}- \frac{1}{2}
	g^{kl}C_{kl}g_{ij}$ is diagonal.  Thus we have

\begin{equation*}
\begin{split}
	\dot b_i & = \frac{b_i}{a_i} b_i -\frac{1}{2} \sum_{k=1}^3
	\frac{b_k}{a_k} b_i + \frac{1}{4}\left( \left(\sum_{k=1}^3
	\frac{b_k}{a_k}\right)^2 - 3\sum_{k=1}^3
	\left(\frac{b_k}{a_k}\right)^2\right) a_i - 	
\\ & \quad - \frac{1}{2} \left(
  	\sum_{k=1}^3\left(\frac{b_k}{a_k}\right)^2 - \frac{1}{2}
	\left(\sum_{k=1}^3 \frac{b_k}{a_k}\right)^2 + \frac{3}{4}
	\left(\left(\sum_{k=1}^3 \frac{b_k}{a_k}\right)^2 - 3\sum_{k=1}^3
	\left(\frac{b_k}{a_k}\right)^2\right) \right) \, a_i
\\ & = \frac{b_i}{a_i} b_i -\frac{1}{2} \sum_{k=1}^3
	\frac{b_k}{a_k} b_i + \frac{1}{8}\left( \left(\sum_{k=1}^3
	\frac{b_k}{a_k}\right)^2 -
	\sum_{k=1}^3\left(\frac{b_k}{a_k}\right)^2 \right) \, a_i.
\end{split}
\end{equation*}

\noindent
	In particular, we have proved \ref{t20ii}.  The eigenvalues of
	$g^{ij} \dot g_{jk}$ are given by $c_{i} = b_i/a_i$.  Expressing
	the system of differential equations with respect to $c_i$ and
	using $\dot c_i = \dot b_i /a_i - (c_i)^2$ we obtain

\begin{equation}	\label{e110}
	\dot c_i = -\frac{1}{2} \sum_{k=1}^3 c_k \, c_i + \frac{1}{8}
	\left(\left(\sum_{k=1}^3 c_k\right)^2 -
	\sum_{k=1}^3\left(c_k\right)^2 \right).
\end{equation}

\noindent
	Since $\Lambda = 0$ Equation (\ref{e95}) implies $64\pi \e =
	\left(\sum_{k=1}^3 c_k\right)^2 - \sum_{k=1}^3\left(c_k\right)^2$.
	Taking the derivative we obtain

\begin{equation*}
	64\pi\dot \e = 2 \left( \sum_{k=1}^3 c_k \sum_{l=1}^3 \dot c_l -
	\sum_{k=1}^3 c_k \dot c_k\right)
\end{equation*}

\noindent
	and inserting Equation (\ref{e110}) gives (with $\alpha :=
	\sum_{k=1}^3 c_k $, $\beta^2 := \sum_{k=1}^3\left(c_k\right)^2$)

\begin{equation*}
\begin{split}
	64\pi\dot \e &= 2 \left( \alpha \sum_{l=1}^3 \left(-\frac{1}{2}
	\alpha\, c_l + \frac{1}{8} \left(\alpha^2 - \beta^2 \right)\right)
	- \sum_{k=1}^3 c_k \left(-\frac{1}{2} \alpha \, c_k +
	\frac{1}{8} \left(\alpha^2-\beta^2 \right)\right) \right)
\\ & =  -\alpha \left( \alpha^2-\beta^2\right) = -64 \pi \alpha \e.
\end{split}
\end{equation*}

\noindent
	On the other hand, Equation (\ref{e70}) reads $\dot \e = -
	\frac{1}{2} \alpha \e$.  Hence we have $\e = 0$ and \ref{t20i} is
	proved.  Assertion \ref{t20iv} is equivalent to
	$\left(\sum_{k=1}^3 c_k\right)^2 - \sum_{k=1}^3\left(c_k\right)^2
	= 0$ and therefore follows from $\e = 0$.  Equation (\ref{e110})
	simplifies to $\dot c_i = -\frac{1}{2} \sum_{k=1}^3 c_k c_i$.
	Thus $\p \left(\sum_{k=1}^3 c_k \right)/ \p t = -\frac{1}{2}\left(
	\sum_{k=1}^3 c_k\right)^2$ which is easily integrated.  Inserting
	the result into our differential equation for $c_i$ we have a
	system of linear, uncoupled differential equations which
	immediately gives \ref{t20iii}.
\end{proof}

\noindent
	The conditions in theorem \ref{t20} together with the usual
	constraint equations ($T_{jt} = 0$ in our case) are necessary but
	need not be sufficient.  On the other hand, there do exist
	non-flat vacuum solutions.  In the next theorem we specialize to a
	case where one can solve the constraint equation easily.

\begin{theorem}	\label{t22}
	Let $(\D, \dg)$ be a 3-dimensional, flat manifold and assume that
	$\hc$ is an analytic, bilinear, symmetric tensor field such that
	there exists exists a Gau\ss ian, orthonormal frame with respect
	to which $\hc$ is diagonal.

	Then there exists a 4-dimensional, discontinuously signature
	changing vacuum spacetime $(M,g)$ such that 
\begin{enumerate}
\item
	$(\D, \dg)$ is the hypersurface of signature change,
\item
	$\hc(\xvec3) = \Ddg(0,\xvec3)$
\item
	$\Dg$ is analytic
\end{enumerate}
if and only if the eigenvalues $\hc_i$ satisfy
$\hc_1\hc_2+\hc_2\hc_3+\hc_3\hc_1 = 0$ and either

\begin{romanlist}
\item
	the $\hc$ are constant with respect to $(\xvec{3})$ or
\item
	$\hc$ depends on only one variable (say, $x^1$) and the components
	$\hc_2$, $\hc_3$ vanish identically.
\end{romanlist}

\noindent
	Moreover, $(M,g)$ is flat if and only if two of the $\hc_i$
	vanish identically.
\end{theorem}  

\begin{proof}
	In this proof we will dismiss Einstein's summation convention if
	the repeated index is $i$.  Our assumptions imply that there
	exists coordinates such that at $t = 0$ we have $g_{ij} =
	\delta_{ij}$ and $c_{ij} = \hc_i\delta_{ij}$.  The (usual)
	constraint equations takes the form

\[
	T_{it} = \Ric_{it}  =\frac{1}{2}g^{jk}
	\left(\p_k\dot{g}_{ij}-\p_i\dot{g}_{jk}\right) = 0
\]

\noindent
	which simplifies to $\p_i \hc_i - \p_i(\hc_1+\hc_2+\hc_3) = 0$.
	Thus there exist positive functions
	$f(x^2,x^3),g(x^1,x^3),h(x^1,x^2)$ and constants
	$\alpha,\beta,\gamma = \pm 1$ such that $\hc_1(\xvec{3}) +
	\hc_2(\xvec{3}) = \gamma (h(x^1,x^2))^2$, $\hc_2(\xvec{3}) +
	\hc_3(\xvec{3}) = \alpha (f(x^2,x^3))^2$, $\hc_3(\xvec{3}) +
	\hc_1(\xvec3) = \beta (g(x^1,x^3))^2$.  Without loss of generality
	we can assume $\alpha = \gamma$.  Now we can express $g(x^1,x^3)$
	in terms of $f(x^2,x^3),h(x^1,x^2)$ using theorem \ref{t20}
	\ref{t20iv}.  We obtain $f = g = h = 0$ or $\alpha = \beta$ and
	$g(x^1,x^3) = f(x^2,x^3) + \delta h(x^1,x^2)$, where $\delta = \pm
	1$.  Hence there exist functions $u(x^1), v(x^2), w(x^3)$ such
	that $f(x^2,x^3) = v(x^2)+w(x^3)$, $g(x^1,x^3) = w(x^3)+ \delta
	u(x^1)$, $h(x^1,x^2) = u(x^1)-\delta v(x^2)$.  With theorem
	\ref{t20} \ref{t20iii} and $c_i = (\ln g_{ij})\spdot \delta_{ij}$
	we obtain for the metric components

\[
	g_{ij}(t,\xvec3) = \delta_{ij} \left(1+ \sum_{k=1}^3\hc_k(\xvec3)
	\, t/2\right)^{\frac{2\hc_i(\xvec3)}{\sum_{k=1}^3\hc_k(\xvec3)}}.
\]

\noindent
	Now it is straightforward to calculate the curvature expressions
	for our candidates of solution\footnote{These calculations have
	been performed using the software package `GRTensorII' for Maple V
	release 3 \cite{musgrave-pollney+94}.  From a purely computational
	point of view it is advantageous to replace the expression
	$\sum_{k=1}^3\hc_k(\xvec3)$ by a general function $L(\xvec3)$.}:

\begin{align*}
\begin{split}
	\frac{\p\DRic_{x^1x^2}}{\p t}(0,\xvec3) &=
		-\frac{1}{2}\alpha\delta \frac{\d v}{\d x^2}(x^2) \frac{\d
		u}{\d x^1}(x^1),
\\ 	\frac{\p\DRic_{x^2x^3}}{\p t}(0,\xvec3) &=
		\frac{1}{2}\alpha\frac{\d v}{\d x^2}(x^2) \frac{\d w}{\d
		x^3}(x^3),
\\ 	\frac{\p\DRic_{x^3x^1}}{\p t}(0,\xvec3) &= \frac{1}{2}\alpha\delta
		\frac{\d w}{\d x^3}(x^3) \frac{\d u}{\d x^1}(x^1).
\end{split}
\end{align*}

\noindent
	Recall that by theorem \ref{t20} \ref{t20v} $\DRic$ must vanish
	for all $t$.  Thus at least two of the functions $u, v, w$ are
	constant.  We will assume that $v$ and $ w$ are constant.  But
	then we have

\[
	\frac{\p\DRic_{x^3x^3}}{\p t}(0,\xvec3) = -\frac{1}{2}\alpha\delta (v+w)
	\frac{\d^2 u}{(\d x^1)^2}(x^1),
\]

\noindent
	whence $v = -w$ or ${\d^2 u}/{(\d x^1)^2}(x^1) = 0$.  If we assume
	${\d^2 u}/{(\d x^1)^2}(x^1) = 0$ then we obtain

\[
	\frac{\p^2\DRic_{x^1x^1}}{\p^2 t}(0,\xvec3) = -\frac{1}{2}\alpha\delta (v+w)
	\left(\frac{\d u}{\d x^1}(x^1)\right)^2.
\]

\noindent
	Thus either $u, v, w$ are all constant or we have $v = -w$ and
	$u(x)$ arbitrary.  Both cases lead to solutions of the vacuum
	equation $\Ric = 0$.  This gives (i) and (ii).  In the case $v =
	-w$ the flat solution is recovered.  In the other case the metric
	is flat if and only if $v = \delta u$ or $v = -w$ or $w = -\delta
	u$. (This can be shown by explicitly calculating the Riemann
	tensor).  By inserting these conditions into $\hc_i$ we conclude
	that $(M,g)$ is flat if and only if two of the $\hc_j$ vanish
	identically.
\end{proof}

\noindent
	We will now consider solutions where $\e$ is continuous but where
	$t \rightarrow \Dg_{ij}$ is only $C^{2-}$, i.e. the one sided
	limits of the second $t$-derivative exist.  These solutions are
	still restricted considerably:

\begin{theorem}	\label{t25}
	Let $(M,g)$ be an m-dimensional, discontinuously signature
	changing spacetime with an adapted dust model. If $t \mapsto \Dg(\xarray)$
	is $C^{1}$ and $\e$ is continuous with respect to
	$(\ttt,\txarray)$, then the surface of signature change $(\D,\dg)$
	satisfies $\ds = 2\Lambda$.

	Conversely, assume that $(\D,\dg)$ is a 3-dimensional, analytic
	Riemannian manifold with $\ds = 2\Lambda$.  If there exists a real
	analytic, bilinear form $\hc$ on $\D$ such that
	$g^{ik}\left(\dnabla_k \hc_{ij}-\dnabla_j\hc_{ik}\right) = 0$,
	then there exists a discontinuously signature changing spacetime
	with an adapted dust model such that
\begin{romanlist}
\item
	$\Dg$ is $C^{2-}$ with respect to $\ttt$ and real analytic with
	respect to $(\txarray)$,
\item
	at $\D$, $\dot g_{ij}(0,\xvec3) = \hc(\xvec3)$ holds,
\item
	$\e$ is continuous with respect to $(\ttt,\txarray)$,
\end{romanlist}
\end{theorem}

\begin{proof}
	The first part follows directly from 	
\[
	8\pi\epsilon =	\frac{\eta}{2}\,\Ds+\frac{1}{8}\left(\left(\tr(\left(
	\Dg\right)\spdot)\right)^2 -\left|\left(
	\Dg\right)\spdot\right|^2\right) -\eta \Lambda.
\]

\noindent
 	For the converse note that for both the Riemannian and the
	Lorentzian region each there exists a unique solution $g_{ij}$
	satisfying the system of equations (\ref{e80}), (\ref{e90}) with
	$\phi = 0$ and $\dot g_{ij}(0,\xvec3) = \hc(\xvec3)$.  By
	assumption the constraints are initially satisfied and as in the
	proof of theorem 2 in \cite{kossowski-kriele-94b} one sees that
	they are then also satisfied everywhere.  Clearly, the solution is
	$C^{2-}$ since it is $C^1$ and obtain by matching of analytic
	solutions.
\end{proof}

\noindent 
	Notice that $\ds = 2\Lambda$ is an {\em additional\/} condition.
	If one replaces $C^1$ by $C^2$ in the above theorem then one
	obtains that $(\D,\dg)$ is flat.  The argument is almost
	identically but employs lemma \ref{l10} for $l = 0$.  On the other
	hand, if one does not assume that $\e$ is continuous then there
	exist as much real analytic solutions as in a purely Lorentzian
	setting.


\subsection{Continuity of the modified energy density $\eta\e$} \label{ss25.20}

Since $\e = T_{\ttt\ttt}$ it is conceivable that one should not assume
smoothness of $\e$ but smoothness of $\tilde\e = \eta\e$.

\begin{theorem}\label{t30}
	Let $(M,g)$ be a 4-dimensional, discontinuously signature changing
	spacetime with an adapted dust model. If $\Dg, \eta\e$ are real
	analytic with respect to $(\ttt,\txvec3)$ and $\e \neq 0$, then
	$g$ is static and the surfaces $\ttt = \text{const}$ have constant
	curvature.

	Conversely, for each 3-dimensional, Riemannian constant curvature
	manifold $(\D,\dg)$ there exists a static, discontinuously signature
	changing dust solution such that $\eta\epsilon$ is analytic and
	$(\D,\dg)$ is the hypersurface of signature change.
\end{theorem}

\begin{proof}
	As in the proof of theorem \ref{t20} we obtain $\DRic =
	2\Lambda\Dg$ for all $t$.  Since $\dim(\D_t) = 3$ the Ricci tensor
	of $(\D_t,\Dg)$ completely determines the Riemann tensor.  Hence
	the hypersurfaces $\D_t$ have constant curvature.  We also have
	Equation (\ref{e95}).  But instead of concluding $\Lambda = 0$ we
	infer $\left(\tr(\left( \Dg\right)\spdot)\right)^2 -\left|\left(
	\Dg\right)\spdot\right|^2 = 0$ for $t = 0$.  Moreover, by
	successive differentiation and analyticity of $\Dg$ we obtain that
	$\left(\tr(\left( \Dg\right)\spdot)\right)^2 -\left|\left(
	\Dg\right)\spdot\right|^2$ vanishes for all $t$.  It follows that
	$8\pi \eta\e = {2}\Lambda$ is constant.  Since $\e \neq 0$ theorem
	\ref{t10} implies $\tr(\left( \Dg\right)\spdot) = 0$ for all $t$
	and therefore $\left|\left( \Dg\right)\spdot\right| = 0 $ for all
	$t$.  But this means that the components $\dot g_{ij}$ vanish and
	hence that $(M,g)$ is static.

	For the converse it is sufficient to check that $\p_\ttt g_{ij} =
	0$ and the condition that $(\D_t,g_{ij})$ has constant curvature
	imply that the Einstein equation is satisfied for an appropriate
	cosmological constant $\Lambda$.
\end{proof}

\noindent
	For completeness, we give also the existence and uniqueness
	theorem for the case where $t \mapsto \Dg(\xarray)$ is a $C^{2-}$-function.

\begin{theorem} \label{t40}
	Let $(M,g)$ be an m-dimensional, discontinuously signature
	changing spacetime with an adapted dust model. If $t \mapsto \Dg(\xarray)$
	is $C^{1}$ and $\eta\e$ is continuous with respect to
	$(\ttt,\txarray)$, then at the hypersurface of signature change
	the second fundamental form $\hc(\xvec3) = \dot g_{ij}(0,\xvec3)$
	satisfies $(\tr(\hc))^2 - |\hc|^2 = 0$.
	
	Conversely, assume that $(\D,\dg)$ is a 3-dimensional, analytic
	Riemannian manifold.  If there exists a real
	analytic, bilinear form $\hc$ on $\D$ such that
	$g^{ik}\left(\dnabla_k \hc_{ij}-\dnabla_j\hc_{ik}\right) = 0$ and
	$(\tr(\hc))^2 - |\hc|^2 = 0$, then there exists a discontinuously
	signature changing spacetime with an adapted dust model such that
\begin{romanlist}
\item
	$\Dg$ is $C^{2-}$ with respect to $\ttt$ and real analytic with
	respect to $(\txarray)$,
\item
	at $\D$, $\dot g_{ij}(0,\xvec3) = \hc(\xvec3)$ holds,
\item
	$\eta\e$ is continuous with respect to $(\ttt,\txarray)$,
\end{romanlist}
\end{theorem}

\begin{proof}
	The proof is analogous to the proof of theorem \ref{t25}. 
\end{proof}

\section{Conclusion}\label{s50}

	It is straight forward to define discontinuously signature
	changing spacetimes as smooth (or even analytic) objects: Writing
	$g = -\eta\d\ttt^2+ g_{ij}(\ttt,\txarray)\d\tx^i\d\tx^j$ one can
	consider the one-parameter family of Riemannian metrics
	$g_{ij}d\tx^i\d\tx^j$.  Weak junction conditions are  implied
	by the requirement that this one-parameter family depends smoothly
	on the parameter $\ttt$ (many authors only demand $C^1$ which is
	also possible).  Requiring the Einstein equation in both the
	Riemannian and the Lorentzian region is then a natural
	generalization of general relativity.  

	This theory can only give rise to explanations if additional
	assumptions are imposed.  Our main assumption in this paper was
	that $(M,g)$ is a dust spacetime with energy density $\e$ such
	that either $\e$ or $\eta\e$ is continuous.  In theorems
	\ref{t20}, \ref{t22} and \ref{t30} we have seen that only for very
	special solutions the 1-parameter family $\{\Dg_{ij}\}$ is
	analytic.  If one assumes that $t \mapsto \Dg_{ij}(\xarray)$ is
	$C^{2-}$, the class of solutions is still more restricted than in
	the purely Lorentzian case.\footnote{Observe that there exist as
	many solutions of low differentiability ($C^{2-}$) with {\em
	discontinuous energy density\/} as in the purely Lorentzian case.
	One can just solve the Riemannian and the Lorentzian part for the
	same analytical initial conditions separately using the theorem of
	Cauchy Kowalewska.}  It follows that the regularity conditions
	imposed by us effectively restrict the space of solutions but
	still leave room for non-trivial spacetimes.  This result should
	be compared with analogous results for other implementations of
	signature change:

\begin{romanlist}
\item
	In the case of smooth signature with strong junction condition the
	hypersurface of signature change must be totally geodesic.  This
	accounts for half the initial conditions for Einstein's equations.
	The existence of a totally geodesic hypersurface is a highly
	non-generic feature and has been used by Hayward \cite{hayward-93a}
	to link smooth signature change to inflation.  It should be noted
	that this interpretation is only possible if the strong energy
	condition is not valid near the hypersurface of signature change.
	In fact, by a slight modification of a singularity theorem of
	Hawking, spacetime would collapse if this energy condition was
	satisfied.  This would clearly be in disagreement with
	observation.
\item
	If one considers discontinuous signature change with strong
	junction conditions, then one obtains different answers according
	to the differentiability conditions one imposes on the spacelike
	metric components $g_{ij}$. If one merely assumes that the
	$g_{ij}$ are $C^1$ but not necessarily $C^2$ then one can recover
	the result from smooth signature change.  If the $g_{ij}$ are
	assumed to be analytic functions and the energy momentum tensor is
	assumed to be smooth then spacetime must be static.  Under these
	conditions, the only solution to the vacuum equation is flat space
	\cite{kossowski-kriele-93b}.
\end{romanlist}

\noindent
	If the theory of discontinuous signature change with weak junction
	conditions is viable then our regular solutions should be of
	special interest.  Unfortunately, assuming the regularity
	conditions of theorem \ref{t20} or theorem \ref{t30}, the
	surviving solutions are static or vacuum and therefore do not seem
	to agree with observation.  While this may be considered as a hint
	against signature change with weak junction conditions, it should
	be kept in mind that we have only considered a very simple
	macroscopic matter model.  It would be interesting to learn
	whether more sophisticated matter models could lead to solutions
	which are physically more realistic.  Theorem \ref{t10} shows that
	are also very restricted, which may be viewed as a preliminary
	result in this direction.  Theorems \ref{t25} and \ref{t40} can be
	easily generalized to scalar field matter models.

	It is interesting to observe that assuming vanishing cosmological
	constant, isotropy, a dust matter model, and continuous energy
	density, only $(k=0)$-Robertson-Walker-spacetimes are compatible
	with signature change under weak regularity conditions (theorem
	\ref{t25}, see also \cite{ellis-sumeruk+92a}).  Adopting signature
	change, we would therefore have a simple explanation of the
	flatness problem which partially motivated inflation.  If (in this
	setting) the assumption of continuous energy density is replaced
	by the condition that $\eta\e$ is continuous, then the
	hypersurface of signature change must be totally geodesic.  This
	would imply that the universe is collapsing in contradiction to
	experience.

	The reader should also keep in mind that smooth and discontinuous
	signature changing dust models with strong junction conditions are
	ruled out by a very different mechanism.  Pressureless dust
	satisfies the strong energy condition and therefore dust
	spacetimes should be collapsing in the Lorentzian region --- in
	contradiction to observation.  Still, in this context it may be
	worthwhile to study alternative matter models which violate the
	energy conditions near the hypersurface of signature change.
	Another way to save signature change with strong junction
	conditions would be to impose signature change in the future
	rather than in the past.
	
	It seems that the weak junction conditions for continuous energy
	density corresponds best to observation if one assumes that matter
	satisfies the strong energy condition near the hypersurface of
	signature change.  However, it should be remarked that this energy
	condition may not be justified, given the early stage of the
	universe's evolution where signature change is supposed to occur.

	Finally, it should be noted that there is still much controversy
	about which implementation of signature change should be
	considered `correct'.  See
	\cite{ellis-sumeruk+92a,hayward-92a,hayward-93-p-b,kossowski-kriele-93b,kriele-martin-95a,carfora-ellis-95a,hellaby-dray-94a,embacher-95a,dray-hellaby-95-p-a,hayward-95a,dray-hellaby-95a},
	for instance.  While in Hayward's papers there have been used many
	harsh words with respect to weak junction conditions, I am not
	aware of any previous work which puts them to the physical test
	examing the {\em consequences\/} of the proposal.

\bigskip

\begin{acknowledgement}
	I would like to thank Franz Embacher and Tevian Dray for
	discussions about the weak junction conditions. 
\end{acknowledgement}

\ifsubmission\pagebreak\fi

\end{document}